\documentclass[12pt,letterpaper]{article}

\usepackage{amsmath}
\usepackage{amsfonts}
\usepackage{amssymb}
\usepackage{graphicx}
\usepackage{fancyvrb}
\usepackage{mathtools}
\usepackage{comment}
\usepackage{cite}
\usepackage{hyperref}
\usepackage{url}

\VerbatimFootnotes
\usepackage[letterpaper,margin=1in]{geometry}

\usepackage{caption}
\usepackage{subcaption}

\begin{document}

\begin{titlepage}

\begin{center}

\vspace*{1.0cm}

{\LARGE\bfseries The s-Rule from Geometry}

\vspace{1.0cm}

{\normalsize
Sebasti\'an Franco$^{a,b\,1}$ and
Diego Rodr\'iguez-G\'omez$^{c,d\,2}$
}

\vspace{0.7cm}

{\normalsize\itshape
$^{a}$ Physics Department, The City College of the CUNY\\
160 Convent Avenue, New York, NY 10031, USA
}

\vspace{0.25cm}

{\normalsize\itshape
$^{b}$ Physics Program and Initiative for the Theoretical Sciences\\
The Graduate School and University Center, The City University of New York\\
365 Fifth Avenue, New York, NY 10016, USA
}

\vspace{0.25cm}

{\normalsize\itshape
$^{c}$ Department of Physics, Universidad de Oviedo\\
C/ Federico Garc\'ia Lorca 18, 33007 Oviedo, Spain
}

\vspace{0.25cm}

{\normalsize\itshape
$^{d}$ Instituto Universitario de Ciencias y Tecnolog\'ias Espaciales
de Asturias (ICTEA)\\
C/ de la Independencia 13, 33004 Oviedo, Spain.
}

\vspace{0.8cm}

{\normalsize\bfseries Abstract}

\vspace{0.3cm}

\begin{minipage}{0.82\textwidth}
\normalsize
String Theory provides powerful tools for constructing and studying five-dimensional Quantum Field Theories, notably through geometric engineering and five-brane webs. Understanding how these two descriptions are related in the presence of 7-branes requires incorporating the s-rule into the geometry. To address this question, we focus on basic three-legged five-brane webs ending on 7-branes. Building on the local M-theory description of five-branes ending on 7-branes, we consider triangular toric diagrams with a single non-primitive edge. We show that the corresponding local complex-structure deformation is well defined on the overlaps with the neighboring toric patches, while the corresponding gluing map remains well behaved as a function of the holomorphic coordinates throughout the neighboring patches precisely when the s-rule is satisfied. At saturation, the geometry is a T-cone and the 7-brane deformation becomes its Q-Gorenstein smoothing. Our results thus identify the s-rule with the obstruction to globalizing the local 7-brane deformation and provide a direct geometric link between T-cones and five-branes ending on 7-branes.
\end{minipage}

\end{center}

\vfill

{\normalsize
\noindent
$^{1}$\texttt{sebastian.franco@ccny.cuny.edu}

\noindent
$^{2}$\texttt{d.rodriguez.gomez@uniovi.es}
}

\end{titlepage}

\tableofcontents

\section{Introduction and conclusions}

String Theory provides powerful tools for engineering Quantum Field Theories whose very existence is far from manifest from conventional field-theoretic constructions. Five-dimensional Superconformal Field Theories (SCFTs) provide a particularly striking example: despite the absence of conventional weakly-coupled Lagrangian descriptions for generic fixed points, large classes of such theories can be defined through String Theory. Two of the most successful constructions are geometric engineering in M-theory on singular Calabi--Yau threefolds \cite{Morrison:1996xf,Douglas:1996xp,Intriligator:1997pq,Closset:2018bjz} and $(p,q)$ 5-brane webs suspended from 7-branes in type IIB String Theory \cite{Aharony:1997bh}. When every external 5-brane ends on a separate 7-brane, the latter may effectively be moved to infinity and the web is related to a toric Calabi--Yau threefold by the standard brane/toric correspondence of Leung and Vafa \cite{Leung:1997tw}. This correspondence, however, captures only a restricted class of 5-brane webs and hence of 5d SCFTs.

A much richer class arises when several 5-branes end on the same 7-brane. Such configurations, which underlie the {\it generalized toric polygons} (GTPs) introduced in \cite{Benini:2009gi} and further developed in \cite{Bourget:2023wlb}, are subject to the s-rule.\footnote{The s-rule was originally formulated in the Hanany--Witten brane construction \cite{Hanany:1996ie} and was subsequently derived and generalized in several related brane and string-junction settings \cite{Bachas:1997sc,Kitao:1998mf,Mikhailov:1998bx,Bergman:1998ej,DeWolfe:1998bi,Iqbal:1998xb}.} The s-rule is a genuinely stringy consistency condition with direct and highly non-trivial consequences for the associated Quantum Field Theories: it constrains which brane configurations are supersymmetric and, consequently, which strongly-coupled fixed points they engineer. At the same time, the presence of the 7-branes prevents the direct toric interpretation of the web. Finding the geometric counterpart of general GTPs, and in particular understanding how the s-rule is encoded geometrically, is therefore an important step towards extending geometric engineering beyond the ordinary toric setting (see also \cite{Alexeev:2024bko} for recent progress on non-toric brane webs and Calabi--Yau threefolds).

Recent progress has begun to bridge this gap. The relation between Hanany--Witten moves and polytope mutations provides a geometric handle on GTPs \cite{Franco:2023flw,Arias-Tamargo:2024fjt}, with polytope mutations admitting a description in terms of flat families of toric varieties \cite{Ilten_2012}, while the local effect of 7-branes has been described in terms of non-toric complex-structure deformations \cite{Bourget:2023wlb,Collinucci:2026kom}. A complementary route starts directly from the brane web. In \cite{Bergman:2020myx}, the extended Coulomb branch was exposed by opening a web as much as supersymmetry allows, decomposing it into irreducible triple junctions saturating the self-intersection bound $\mathcal{I}=-2$. Such junctions have no hidden faces and therefore cannot be opened further. A characteristic feature of these elementary junctions is that, when regarded in isolation, an appropriate 7-brane crossing turns the three-legged junction into a configuration in which one 7-brane is detached.

The {\it 5d tangram} construction \cite{CarrenoBolla:2024fxy} identified the geometric counterpart of precisely these rigid $\mathcal{I}=-2$ junctions. Remarkably, they are the T-cones introduced in the mathematical literature: special cyclic quotient singularities admitting Q-Gorenstein smoothings \cite{Shepherd1988}. This led to a geometric description of the extended Coulomb branch of a GTP in terms of tessellations by T-cones (and, more generally, their locked superpositions). The identification brought together two a priori rather different notions: on the brane side, an elementary triple junction singled out by the supersymmetry bound; on the geometric side, an orbifold singled out by the existence of a Q-Gorenstein smoothing. In the tangram picture the smoothing was identified physically with the 7-brane operation that converts the toric configuration with $p$ separate external 5-branes into the GTP in which those $p$ 5-branes end on a common 7-brane.

There is a further reason why T-cones are natural elementary objects in this problem. An external leg of a GTP, including its 7-brane, can itself be viewed as a T-cone: the $p$ 5-branes ending on the 7-brane, together with the two neighboring 5-branes, form an $\mathcal{I}=-2$ three-legged junction with no internal face. This perspective also makes contact with Hanany--Witten transitions. As argued in \cite{Arias-Tamargo:2024fjt,CarrenoBolla:2024fxy}, crossing a 7-brane through a web is mapped geometrically to a polytope mutation. In the T-cone language, the elementary operation consists of removing the T-cone associated with the external leg from one side of the polygon and attaching its mutated image on the other side. The branes created in the Hanany--Witten process are encoded in this mutation. T-cones therefore play a dual role: they are the elementary pieces from which GTPs are tessellated and the elementary objects acted on by Hanany--Witten transitions.

This structure also appears to leave a non-trivial imprint on the BPS quiver and its brane-tiling description; for the relation between 5d SCFTs, BPS quivers and brane tilings see also \cite{Closset:2019juk}. In \cite{Franco:2026bie}, it was shown that, once an origin is chosen, assigning suitable scalings to the fields of the ordinary toric quiver produces a Hilbert series equal to the Ehrhart series of the dual polytope, and hence invariant under the mutations associated with Hanany--Witten transitions. More recently, \cite{Franco:2026hiz} proposed a dimer avatar of the operation that defines a GTP: the $p$ parallel 5-branes ending on a common 7-brane correspond to parallel zig-zag paths bounding $\mathcal N=2$ strips, and bringing the 5-branes onto the same 7-brane translates into condensing these strips. The corresponding gauge groups are driven to infinite coupling and subsequently confine. Since the strips that condense are precisely those associated with the T-cone, this gives yet another manifestation of its elementary role. It also suggests an intriguing relation between the two descriptions: the mutation-compatible field scalings visible already in the original toric BPS quiver may be the ultraviolet algebraic avatar of the strongly-coupled infrared dynamics produced by T-cone strip condensation \cite{Franco:2026bie,Franco:2026hiz}. We will not pursue this connection here, but the geometric picture developed below provides a complementary avatar of the same operation.

These observations sharpen the relation between the physical and mathematical appearances of T-cones. On the brane-web side, opening webs until no hidden face remains naturally leads to irreducible $\mathcal{I}=-2$ junctions \cite{Bergman:2020myx}. The tangram construction relates these junctions to the T-cones of the mathematical literature, whose key property for the present purposes is the existence of a Q-Gorenstein smoothing \cite{Shepherd1988,CarrenoBolla:2024fxy}. This brings together two seemingly different structures, but leaves open a basic question: why should the mathematical smoothing know about the physical operation of terminating several 5-branes on a common 7-brane?

The purpose of this paper is to close this circle directly, in the simplest non-trivial setting. To do so, we apply the geometric framework recently developed in \cite{Collinucci:2026kom} to the T-cones introduced in \cite{CarrenoBolla:2024fxy}, thereby making the relation between the two descriptions explicit. Extending this analysis to more general webs is a natural and interesting problem, which we leave for future work. We start from the orbifold geometry associated with a three-legged web and construct the deformation induced by forcing the $p$ 5-branes of one external leg to end on a common 7-brane. From the supersymmetric viewpoint, the natural requirement is that the corresponding gluing map be expressible entirely in terms of the holomorphic coordinates, without involving their complex conjugates. The non-trivial geometric requirement is then that this map remain well defined and finite throughout the neighboring patches. We show that this is possible precisely when the s-rule is satisfied. At saturation, the geometry is exactly a T-cone and the deformation is exactly its Q-Gorenstein smoothing. In other words, for the T-cone the abstract mathematical smoothing has a concrete brane interpretation: it is the geometric operation that takes the $p$ would-be separate external legs of the orbifold and makes them end on the same 7-brane. The resulting web is precisely the rigid three-legged configuration from which the T-cone entered the brane-web story in the first place. Thus the $\mathcal{I}=-2$ junctions arising from opening webs, the T-cones used to tessellate GTPs, the T-cones moved by Hanany--Witten transitions, and the cyclic quotient singularities singled out by Q-Gorenstein smoothability are different facets of the same structure. This provides a direct geometric origin for the basic ingredients of the tangram construction.

The rest of the paper is organized as follows. In section \ref{sec:localleg} we review the local 5-brane-7-brane system and its M-theory description, emphasizing the complex-structure deformation associated with terminating several 5-branes on a common 7-brane. In section \ref{sec:s-rule} we embed this local deformation into a three-legged toric Calabi--Yau and show that requiring the gluing map to remain well behaved as a function of the holomorphic coordinates throughout the neighboring patches reproduces the s-rule. Section \ref{sec:Tcones} specializes to the saturation case and shows how the T-cone is singled out, with the 7-brane deformation becoming its Q-Gorenstein smoothing. We have kept the main text focused on the geometric mechanism and its physical interpretation, relegating technical details to the appendices: appendix \ref{app:toric} collects the toric patches and transition functions, appendix \ref{app:allorders} establishes the all-orders scaling of the local formal gluing map, and appendix \ref{app:qgorenstein} gives the detailed construction of the Q-Gorenstein smoothing.

\section{The local 5-/7-brane geometry}\label{sec:localleg}

A leg of a $(p,q)$ web ending on a 7-brane is locally described by a collection of D5-branes ending on a D7-brane. We briefly review this local system and its M-theory description, following the geometric formulation used in \cite{Collinucci:2026kom}. Consider the following D5--D7 configuration in type IIB String Theory:

\begin{equation}
\begin{array}{c | c c c c c c c c c c}
& 0 & 1 & 2 & 3 & 4 & 5 & 6 & 7 & 8 & 9\\  \hline
D5 & \times & \times & \times & \times & \times &\times  & -  &-  &  - & - \\
D7 & \times & \times & \times & \times &\times & - & - & \times &\times  & \times
\end{array}
\end{equation}

Taking $x^9$ to be compact and performing T-duality along it gives the type IIA configuration

\begin{equation}
\begin{array}{c | c c c c c c c c c c}
& 0 & 1 & 2 & 3 & 4 & 5 & 6 & 7 & 8 & 9\\  \hline
D6_5 & \times & \times & \times & \times & \times &\times  & -  &-  &  - & \times \\
D6_7 & \times & \times & \times & \times &\times & - & - & \times &\times  & -
\end{array}
\end{equation}

From the viewpoint of the $D6_7$-brane, it is natural to introduce the worldvolume holomorphic coordinate $z=x^7+ix^8$. In turn, the transverse fields $x^5$ and $x^9$ would be naturally combined into the holomorphic quantity $W=x^5+ix^9$, with the SUSY condition given by $W=f(z)$, with $f(z)$ holomorphic. However, since $x^9$ is a circle, it is natural to consider the transverse field $w_{\pm}=e^{\pm W}$. Then, the supersymmetry conditions amount to $w_+=f(z)$. 

To motivate the local profile, temporarily consider the $D6_5$--$D6_7$ system in $\mathbb{R}^{1,9}$. We may then work directly with $W$, which away from the endpoints satisfies $\partial \bar{\partial}W=0$. However, at the points where the $D6_5$ end/intersect the $D6_7$ the LHS develops a $\delta$ function. Thus, what we have to solve is

\begin{equation}
\partial \bar{\partial} W=\pm\delta(z-z_i)\,,
\end{equation}
where $\pm$ depends on whether the branes go to the left or to the right. The solution is

\begin{equation}
W=\pm \log(z-z_i)\,,
\end{equation}
In general

\begin{equation}
W=\sum_{i=1}^{L} \log(z-z_i) - \sum_{i=1}^R \log(z-z_i)\,.
\end{equation}
Thus, in terms of the $w_+$ coordinate

\begin{equation}
w_+= \frac{\prod_{i=1}^L\,(z-z_i)}{\prod_i^R (z-z_i)}\,.
\end{equation}

Suppose now that we only have $p$ branes to the left, and that they all are at the same point. To be clear, this situation corresponds in the original IIB system to $p$ D5-branes ending on one D7-brane, and the further requirement that they end at the same place is motivated by the fact that we want to use this system as a leg of a 5-brane web. Then $w_+=z^p$, which we can write as $\Delta=z^p-w_+=0$. Thus, uplifting to M-theory, where we denote $z\rightarrow Z$ and $w_+\rightarrow S$, this corresponds to

\begin{equation}
\label{patch}
XY=Z^p-S\,.
\end{equation}
Equation \eqref{patch} will be key in the following, as it corresponds to the M-theory avatar of a leg of a $(p,q)$ web: $p$ 5-branes ending on a single 7-brane.

\section{The s-rule for triangles}\label{sec:s-rule}

Consider a toric geometry with a triangular toric diagram with vertices

\begin{equation}
\label{toricvertices}
v_1=(0,0,1)\,,\qquad v_2=(p,0,1)\,,\qquad v_3=(q,m,1)\,.
\end{equation}
The three edges of the triangle define three toric patches. Besides the multiplicity $p$ of the lower edge, it is convenient to introduce

\begin{equation}
d={\rm gcd}(q,m)\,,\qquad e={\rm gcd}(p-q,m)\,,
\end{equation}
which are the lattice multiplicities of the left and right edges, respectively. The corresponding local geometries are

\begin{equation}
\mathcal{U}_b\simeq \frac{\mathbb{C}^2}{\mathbb{Z}_p}\times\mathbb{C}^{\star}\,,
\qquad
\mathcal{U}_l\simeq \frac{\mathbb{C}^2}{\mathbb{Z}_d}\times\mathbb{C}^{\star}\,,
\qquad
\mathcal{U}_r\simeq \frac{\mathbb{C}^2}{\mathbb{Z}_e}\times\mathbb{C}^{\star}\,.
\end{equation}
A derivation of these patches, together with the complete set of transition functions, is given in appendix \ref{app:toric}.

\begin{figure}[ht!]
\centering
\includegraphics[width=4.5cm]{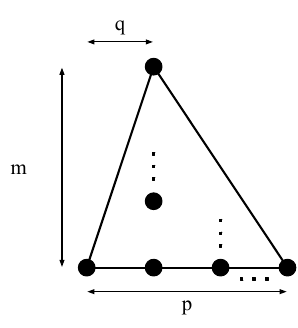}
\caption{The basic toric diagram for the vertices in \eqref{toricvertices} and $d=e=1$.}
\label{basic_toric}
\end{figure}

We are interested in triangles for which the lower edge is the only non-primitive one. We therefore set

\begin{equation}
d=e=1\,.
\end{equation}
The neighboring left and right patches are then smooth, $\mathcal{U}_{l,r}\simeq \mathbb{C}^{2}\times\mathbb{C}^{\star}$, while the lower patch contains the non-trivial
$\mathbb{C}^{2}/\mathbb{Z}_{p}$ singularity. The resulting toric diagram is schematically shown in Figure \ref{basic_toric}.

%=================================================================
%=================================================================

Following our previous discussion, in order to end the $p$ external legs dual to the $\mathbb{C}^{2}/\mathbb{Z}_{p}$ lower patch we deform the equation as 

\begin{equation}
X_bY_b=Z_b^p+t\,S_b\,.
\end{equation}
This deformation translates to the neighboring patches through the transition functions. For later use, we only need the bottom--left and bottom--right transition functions. In the notation derived in appendix \ref{app:toric}, they imply that the deformation of the lower patch is described on the neighboring patches by

\begin{equation}
\label{deformed-neighbours}
X_l^pY_l^p=Z_l^p+t\,\frac{Y_l^m}{S_l^r}\,, \qquad X_r^pY_r^p=Z_r^p+t\,\frac{Y_r^m}{S_r^u}\,.
\end{equation}

Moreover, on the bottom--left and bottom--right overlaps one has, respectively,

\begin{equation}
Y_l\neq 0\,,\qquad Y_r\neq 0\,.
\end{equation}

The two neighboring patches therefore lead to identical gluing problems, and in what follows we concentrate on the left patch. The right patch follows \textit{mutatis mutandis}.

\subsection{The s-rule}

We have seen how the deformation introduced in the bottom patch translates to the neighboring patches through the transition functions. Yet, since the local geometry of the neighboring patches is unchanged, consistent gluing requires the induced deformation to be removable by a change of coordinates which is a function of the holomorphic coordinates. From the supersymmetric viewpoint, the natural requirement is that the gluing map be expressible in terms of the holomorphic coordinates, without involving their complex conjugates; the additional question relevant here is whether this map remains well defined and finite throughout the full neighboring patch. Thus, there must be an invertible change of variables $(X_l,Y_l,Z_l,S_l)\rightarrow (\mathcal{X}_l,\mathcal{Y}_l,\mathcal{Z}_l,\mathcal{S}_l)$ which reduces to the identity for $t=0$ and maps \eqref{deformed-neighbours} to

\begin{equation}
\mathcal{X}_l^p\mathcal{Y}_l^p=\mathcal{Z}_l^p\,,
\end{equation}
where the relevant branch is equivalent to $\mathcal{X}_l\mathcal{Y}_l=\mathcal{Z}_l$ as we explain below.

Let us stress that in the bottom--left overlap one has $Y_l\neq0$. Consequently, a coordinate transformation containing negative powers of $Y_l$ can still define a perfectly good function of the holomorphic coordinates on the overlap and hence a legitimate gluing there. The non-trivial question is whether this transition map remains well behaved---that is, finite---when continued to the full left patch, which includes the locus $Y_l=0$. As we will now show, this is possible precisely when

\begin{equation}
m\geq p\,,
\end{equation}
reproducing the s-rule. The analysis of the right patch is identical.

\subsubsection{Warm-up: $p=2$}

Let us first illustrate the gluing problem in the simplest case, $p=2$. The equation induced on the left patch is
\begin{equation}
X_l^2Y_l^2=Z_l^2+t\,\frac{Y_l^m}{S_l^r}\,.
\label{eq:p2left}
\end{equation}
Consider the change of coordinates
\begin{equation}
(X_l,Y_l,Z_l,S_l)\rightarrow \left( \mathcal{X}_l+t\,\frac{\mathcal{Y}_l^{m-1}}{2\mathcal{S}_l^r(\mathcal{X}_l\mathcal{Y}_l-\mathcal{Z}_l)}, \mathcal{Y}_l, \mathcal{Z}_l+t\,\frac{\mathcal{Y}_l^m}{2\mathcal{S}_l^r(\mathcal{X}_l\mathcal{Y}_l-\mathcal{Z}_l)}, \mathcal{S}_l\right)\,.
\label{eq:p2change}
\end{equation}
In these coordinates, \eqref{eq:p2left} becomes
\begin{equation}
\mathcal{X}_l^2\mathcal{Y}_l^2=\mathcal{Z}_l^2\,,
\end{equation}
Equation (19) has two branches, $\mathcal{X}_l\mathcal{Y}_l=\pm\mathcal{Z}_l$. For the change of coordinates \eqref{eq:p2change}, one has $X_lY_l-Z_l=\mathcal{X}_l\mathcal{Y}_l-\mathcal{Z}_l$. Since the left-hand side cannot vanish on the bottom--left overlap, the branch $\mathcal{X}_l\mathcal{Y}_l=\mathcal{Z}_l$ is not reached there. The transformation therefore selects $\mathcal{X}_l\mathcal{Y}_l=-\mathcal{Z}_l$. Obviously, this is equivalent to the undeformed local geometry $X_lY_l=Z_l$ by the trivial redefinition $\mathcal{Z}_l\rightarrow-\mathcal{Z}_l$. In turn, the inverse transformation is
\begin{equation}
(\mathcal{X}_l,\mathcal{Y}_l,\mathcal{Z}_l,\mathcal{S}_l)\rightarrow \left( X_l-t\,\frac{Y_l^{m-1}}{2S_l^r(X_lY_l-Z_l)}, Y_l, Z_l-t\,\frac{Y_l^m}{2S_l^r(X_lY_l-Z_l)}, S_l\right)\,.
\label{eq:p2inverse}
\end{equation}

We now ask whether the map extends to the whole left patch. The only obstruction is the locus $X_lY_l=Z_l$. On the bottom--left overlap this locus is never reached: substituting $X_lY_l=Z_l$ into \eqref{eq:p2left} gives $t\,Y_l^m/S_l^r=0$, and hence $Y_l=0$, whereas the overlap is defined by $Y_l\neq0$. Thus \eqref{eq:p2change} is a perfectly good invertible transition map, depending only on the holomorphic coordinates, on the overlap. The question is whether it remains well defined and finite when continued to the full left patch, where $Y_l=0$ is allowed.

To study the extension, approach the potentially singular locus by writing
\begin{equation}
X_lY_l=Z_l+\epsilon\,,\qquad \epsilon\rightarrow0\,.
\end{equation}
Equation \eqref{eq:p2left} then gives
\begin{equation}
\epsilon^2+2Z_l\epsilon -t\,\frac{Y_l^m}{S_l^r}=0\,,
\label{eq:p2epsilon}
\end{equation}
up to the irrelevant sign convention for $t$. 

At a point of the deformed geometry at which $\epsilon\to0$, one must in fact also have $Z_l\to0$. Indeed, setting $\epsilon=0$ in \eqref{eq:p2epsilon} first gives $Y_l=0$, and then $X_lY_l=Z_l$ implies $Z_l=0$. Rather than assuming a particular relative scaling between $Z_l$ and $\epsilon$, the behavior of the shifts can be determined exactly on the deformed geometry. Using \eqref{eq:p2epsilon}, one finds
\begin{equation}
\frac{tY_l^{m-1}}{S_l^r\epsilon} =\frac{\epsilon+2Z_l}{Y_l} =X_l+\frac{Z_l}{Y_l}\,,
\label{eq:p2exactshift}
\end{equation}
where in the last equality we used $\epsilon=X_lY_l-Z_l$. Therefore the shifts appearing in \eqref{eq:p2inverse} can be written as
\begin{equation}
\delta X_l=-\frac{1}{2}\left(X_l+\frac{Z_l}{Y_l}\right)\,, \qquad \delta Z_l=-\frac{1}{2}\left(X_lY_l+Z_l\right)\,.
\label{eq:p2exactshifts}
\end{equation}
The second shift is manifestly finite at every finite point of the left patch. Thus the only possible obstruction is the ratio $Z_l/Y_l$. Dividing \eqref{eq:p2left} by $Y_l^2$ on the bottom--left overlap gives
\begin{equation}
\left(\frac{Z_l}{Y_l}\right)^2 =X_l^2-t\,\frac{Y_l^{m-2}}{S_l^r}\,.
\label{eq:p2ratio}
\end{equation}
Since $S_l$ is non-vanishing and $X_l$ is finite throughout the left patch, the ratio $Z_l/Y_l$ remains finite as $Y_l\to0$ provided
\begin{equation}
m\geq2\,.
\end{equation}
For $m<2$, instead, the second term on the right-hand side of \eqref{eq:p2ratio} diverges as $Y_l\to0$, and generically $Z_l/Y_l$ diverges with it. The shift $\delta X_l$ therefore fails to remain finite. Hence, for $p=2$, the deformation always defines an invertible gluing map, depending only on the holomorphic coordinates, across the bottom--left overlap, but this map remains well behaved throughout the full left patch precisely when $m\geq2$. This is the s-rule in this simplest case.

It is useful to recast this exact result in a simpler scaling language. The preceding argument did not assume any relation between $Z_l$ and $\epsilon$, and therefore already controls all possible approaches to the boundary. In particular, \eqref{eq:p2ratio} shows that the strongest possible singular behavior is set by the deformation term,
\begin{equation}
\frac{Z_l}{Y_l}\sim Y_l^{m/2-1}\,.
\label{eq:p2worst}
\end{equation}
Equivalently, the potentially most singular regime has $Z_l$ and $\epsilon=X_lY_l-Z_l$ of order $Y_l^{m/2}$. This same exponent can be isolated directly from \eqref{eq:p2epsilon} by considering the regime in which the $\epsilon^2$ term controls the balance, taking $Z_l=0$ as a convenient representative,
\begin{equation}
\epsilon^2\sim Y_l^m\,, \qquad \epsilon\sim Y_l^{m/2}\,.
\label{eq:p2worst-epsilon}
\end{equation}
The point of \eqref{eq:p2worst-epsilon} is not that every approach to $Y_l=0$ obeys this scaling: different relative scalings of $Z_l$ and $\epsilon$ are certainly possible.\footnote{For example, suppose that $Z_l\sim Y_l^\alpha$ and that $Z_l$ dominates over $\epsilon$. Then \eqref{eq:p2epsilon} gives $2Z_l\epsilon\sim Y_l^m$, and hence $\epsilon\sim Y_l^{m-\alpha}$. Since $X_lY_l=Z_l+\epsilon$ and $X_l$ remains finite, such an approach to the boundary necessarily has $\alpha\geq1$; moreover, the assumption $Z_l\gg\epsilon$ implies $\alpha<m/2$. The potentially singular shift then behaves as $\delta X_l\sim Y_l^{m-1}/\epsilon\sim Y_l^{\alpha-1}$, which is finite and is milder than the maximally singular behavior $Y_l^{m/2-1}$ isolated in \eqref{eq:p2worst}. The complementary regime $\epsilon\gtrsim Z_l$ is precisely the one whose limiting balance gives \eqref{eq:p2worst-epsilon}.} Rather, the exact analysis above shows that \eqref{eq:p2worst-epsilon} captures the worst possible power governing the potentially singular shift. All other balances are no more singular. Thus, for the purpose of testing finiteness, it is sufficient to focus on this maximally singular scaling. We will use this shortcut in the generic-$p$ analysis below.

\subsubsection{Generic $p$}

We now turn to generic $p$. The equation induced on the left patch is
\begin{equation}
\label{eq:genericp}
X_l^pY_l^p=Z_l^p+t\,C\,, \qquad C=\frac{Y_l^m}{S_l^r}\,.
\end{equation}
As in the $p=2$ example, we look for an invertible change of coordinates, depending only on the holomorphic coordinates and connected continuously to the identity. It is sufficient to keep $Y_l$ and $S_l$ fixed and write
\begin{equation}
\label{eq:genericchange}
X_l\rightarrow X_l+A\,, \qquad Z_l\rightarrow Z_l+B\,, 
\end{equation}
where $A$ and $B$ are formal power series in $t$,
\begin{equation}
A=tA_1+t^2A_2+\cdots\,, \qquad B=tB_1+t^2B_2+\cdots\,.
\end{equation}
We determine these functions order by order by requiring that \eqref{eq:genericp} be mapped to
\begin{equation}
X_l^pY_l^p=Z_l^p\,.
\end{equation}
As before, this equation has $p$ branches, $X_lY_l=\omega^k Z_l$, with $\omega^p=1$.\footnote{The coordinate transformation selects one of these branches with $k\ne 0$. The particular choice of branch is immaterial for the geometry, since the different branches are related by a constant phase redefinition of $Z_l$.} At first order in $t$ one finds
\begin{equation}
A_1= \frac{1}{p}\, \frac{C}{Y_l}\, \frac{1}{X_l^{p-1}Y_l^{p-1}-Z_l^{p-1}}\,, \qquad B_1=A_1Y_l\,.
\label{eq:A1}
\end{equation}
For generic $p$, \eqref{eq:A1} should be regarded as a local representative of the coordinate redefinition on the overlap. Unlike for $p=2$, for generic $p$ the denominator can vanish on additional loci corresponding to special relative phases between $X_lY_l$ and $Z_l$. These are artifacts of this particular representative rather than of the deformation itself. We will use \eqref{eq:A1} only to extract the local scaling of the potentially singular shift as the boundary $Y_l=0$ is approached. Reassuringly, in section \ref{sec:Tcones} we will see that this local scaling diagnosis gives precisely the condition obtained directly from the global geometry. To determine the relevant scaling, write
\begin{equation}
X_lY_l=Z_l+\epsilon\,, \qquad \epsilon\rightarrow0\,.
\end{equation}
On the deformed geometry,
\begin{equation}
(Z_l+\epsilon)^p-Z_l^p =t\,\frac{Y_l^m}{S_l^r}\,.
\label{eq:generic-epsilon}
\end{equation}
At the locus where $\epsilon\to0$ one also has $Z_l\to0$: setting $\epsilon=0$ in \eqref{eq:generic-epsilon} gives $Y_l=0$, and then $X_lY_l=Z_l$ implies $Z_l=0$. Motivated by the exact $p=2$ analysis above, the potentially most singular scaling is
\begin{equation}
\epsilon^p\sim Y_l^m\,, \qquad\Longrightarrow\qquad \epsilon\sim Y_l^{m/p}\,.
\label{eq:generic-scaling}
\end{equation}
Using
\begin{equation}
X_l^{p-1}Y_l^{p-1}-Z_l^{p-1} =(Z_l+\epsilon)^{p-1}-Z_l^{p-1}\,,
\end{equation}
one sees that near the relevant locus the first-order shifts scale as
\begin{equation}
A_1\sim Y_l^{m/p-1}\,, \qquad B_1\sim Y_l^{m/p}\,.
\end{equation}
Hence the potentially dangerous shift is the one in $X_l$, and its finiteness requires
\begin{equation}
m\geq p\,.
\label{eq:srulegeneric}
\end{equation}

The same condition persists at higher orders. For example, at second order one obtains
\begin{equation}
A_2= -\frac{p-1}{2p^2}\, \frac{C^2}{Y_l}\, \frac{X_l^{p-2}Y_l^{p-2}-Z_l^{p-2}} {\left(X_l^{p-1}Y_l^{p-1}-Z_l^{p-1}\right)^3}\,, \qquad B_2=A_2Y_l\,,
\label{eq:A2}
\end{equation}
and the same scaling analysis gives again the bound $m\geq p$. As shown in appendix \ref{app:allorders}, the same potentially singular scaling persists order by order for this local formal representative. Thus the patchwise analysis diagnoses the bound
\begin{equation}
m\geq p\,,
\end{equation}
which is exactly the s-rule bound. The right patch gives the same condition. In the next section we will see directly from the global geometry that this condition is not merely a feature of the local representative: it is precisely the condition for the 7-brane deformation itself to extend to the ambient geometry. The saturating case, $m=p$, corresponds to the T-cones and will be discussed next.

\section{T-cones}\label{sec:Tcones}

The patchwise analysis of the previous section diagnoses the s-rule through the behavior of local representatives of the gluing map as the boundary of the neighboring patches is approached. For triangular toric diagrams, the extension problem admits a direct global description, which establishes the criterion invariantly and makes manifest the special role of the saturating case $m=p$.

Since the toric diagram is a triangle, the corresponding affine toric Calabi--Yau can be described as a finite abelian quotient of $\mathbb{C}^3$. Let $N'\subset N\simeq\mathbb{Z}^3$ be the sublattice generated by the three rays $v_i$. The quotient group is $G=N/N'$ (see, for example, Chapter 3, in particular Section 3.3, of \cite{CoxLittleSchenck}). The order of $G$ can be read off from the matrix with columns $v_i$. It reads
\begin{equation}
V=\begin{pmatrix}
0 & p & q\\
0 & 0 & m\\
1 & 1 & 1
\end{pmatrix}\,,
\end{equation}
and its determinant is $|\det V|=pm$. For $m=p$, this is the T-cone orbifold of \cite{CarrenoBolla:2024fxy}.\footnote{Indeed, at $m=p$ the standing assumption $e=\gcd(p-q,p)=1$ implies $\gcd(p,q)=1$. The class of $e_1+e_2$ has coefficients $(-2p+q,p-q,p)/p^2$ in the basis $\{v_i\}$, hence order $p^2=|G|$; thus $G\simeq\mathbb Z_{p^2}$, with generator $(z_1,z_2,z_3)\mapsto(\xi^{-2p+q}z_1,\xi^{p-q}z_2,\xi^p z_3)$, $\xi^{p^2}=1$.} The subgroup relevant for the lower edge can be identified directly. The lattice vector $e_1=(1,0,0)$ satisfies
\begin{equation}
e_1=-\frac{1}{p}v_1+\frac{1}{p}v_2\,.
\end{equation}
Its class in $N/N'$ therefore has order $p$ and generates a $\mathbb{Z}_p$ subgroup acting as
\begin{equation}
(z_1,z_2,z_3)\sim \left( \omega^{-1}z_1,\, \omega z_2,\,z_3 \right)\,,\qquad \omega^p=1\,.
\label{eq:standardZp}
\end{equation}
This is precisely the standard $\mathbb{Z}_p$ action associated with the lower edge. A convenient set of invariant coordinates is therefore
\begin{equation}
A=z_2^p\,,\qquad B=z_1^p\,,\qquad C=z_1z_2\,,\qquad D=z_3\,,
\end{equation}
subject to
\begin{equation}
\label{eq:ABC}
AB=C^p\,.
\end{equation}
Using appendix \ref{app:toric}, in terms of these variables, the coordinates of the lower-edge patch are
\begin{equation}
\label{eq:bottomglobal}
X_b=D^qA\,,\qquad Y_b=D^{p-q}B\,,\qquad Z_b=DC\,,\qquad S_b=D^m\,.
\end{equation}
Substituting \eqref{eq:bottomglobal} into the deformed lower-patch equation $X_bY_b=Z_b^p+tS_b$ gives
\begin{equation}
D^p\left(AB-C^p-t\,D^{m-p}\right)=0\,.
\end{equation}
The lower patch has $S_b=D^m\neq0$, and therefore $D\neq0$ there. We may thus divide by $D^p$ and obtain
\begin{equation}
\label{eq:globaldeformation}
AB-C^p-t\,D^{m-p}=0\,.
\end{equation}
This form makes the extension problem completely transparent. The lower patch only probes $D\neq0$, while extending the deformation to the ambient geometry requires \eqref{eq:globaldeformation} to remain holomorphic at $D=0$. This happens if and only if
\begin{equation}
m\geq p\,,
\end{equation}
in precise agreement with the patchwise gluing analysis.

Equation \eqref{eq:globaldeformation} also singles out the saturating case. For $m>p$, the deformation term vanishes at $D=0$ and the fiber there remains
\begin{equation}
AB-C^p=0\,,
\end{equation}
so the $\mathbb{C}^2/\mathbb{Z}_p$ singularity persists. By contrast, for $m=p$ one obtains
\begin{equation}
\label{eq:localdeformation}
AB-C^p-t=0\,,
\end{equation}
which completely smooths the singular fiber. The smoothing is in fact Q-Gorenstein. As we show explicitly in appendix \ref{app:qgorenstein}, the T-cone can be written as a $\mathbb Z_p$ quotient of $\mathbb C\times\mathbb C^2/\mathbb Z_p$, and the deformation \eqref{eq:localdeformation} is compatible with this quotient. Moreover, the residual $\mathbb Z_p$ action leaves invariant the holomorphic three-form of the deformed geometry, so that the relative canonical divisor of the resulting family is Cartier. Thus the T-cones are precisely the configurations that saturate the s-rule bound, and their distinguished geometric property is that the 7-brane deformation extends to their Q-Gorenstein smoothing.

%======================================================
\subsection{Relation to other geometric approaches}
%======================================================

It is interesting to compare our results with the geometric approach to 5-brane webs with 7-branes developed in~\cite{Alexeev:2024bko}. The starting points and perspectives of the two constructions are rather different. Reference~\cite{Alexeev:2024bko} develops a global algebro-geometric description of webs with 7-branes in terms of log Calabi--Yau surfaces, with the broader goal of characterizing consistent, possibly non-toric, webs and constructing the corresponding Calabi--Yau threefolds. Here, instead, we start directly from the local effect of a 7-brane on a toric Calabi--Yau and ask whether the resulting deformation can be extended in a well-defined manner beyond the patch in which it is initially defined. Remarkably, this simple extension problem reproduces the s-rule, $m\geq p$.  Our analysis therefore provides a complementary perspective on this geometry, as it shows how the T-cone is geometrically singled out as the configuration that saturates the regular-extension bound imposed by the 7-brane.

\section*{Acknowledgements}

S.F. is supported by the U.S. National Science Foundation grant PHY-2412479. D.R.G. is supported in part by the Spanish national grant NAC-ES-PUB-ASV-2025 PID2024-161500NB-I00. S.F. would like to thank Cumrun Vafa for useful comments and Harvard University for hospitality during the completion of this work. We acknowledge the use of ChatGPT (OpenAI) and Claude (Anthropic) for technical discussions and checks, as well as for assistance in improving the grammar and language of the manuscript. The authors take full responsibility for the content and conclusions of the work.

\appendix

\section{Toric patches and transition functions} \label{app:toric}

For completeness, we collect here the derivation of the three edge patches associated with the triangular toric diagram

\begin{equation}
v_1=(0,0,1)\,,\qquad v_2=(p,0,1)\,,\qquad v_3=(q,m,1)\,,
\end{equation}
together with the transition functions relating them. The construction is identical for the three edges, so it is useful to present them in parallel.

\begin{itemize}

\item \textbf{Lower edge}: The lower edge is generated by $\{v_1,v_2\}$. An element of the dual lattice, $m=(a,b,c)\in\mathbb{Z}^3$, belongs to the dual cone when $\langle m|v_i\rangle\geq0$. For this edge it is convenient to introduce

\begin{equation}
V=(\alpha,\beta)=(c,pa+c)\,,
\end{equation}
for which $\alpha-\beta\in p\mathbb{Z}$. The semigroup is therefore generated by $(1,1)$, $(0,p)$ and $(p,0)$. These vectors lift to

\begin{equation}
m_1=(0,0,1)\,,\qquad m_2=(1,0,0)\,,\qquad m_3=(-1,0,p)\,.
\end{equation}
The map $(a,b,c)\mapsto(\alpha,\beta)$ has a one-dimensional kernel, generated by

\begin{equation}
m_4=(0,1,0)\,,
\end{equation}
which accounts for the additional $\mathbb{C}^{\star}$ direction. The relevant pairings are

\begin{align}
&\langle m_1|v_1\rangle=1\,,\qquad
 \langle m_1|v_2\rangle=1\,,\qquad
 \langle m_1|v_3\rangle=1\,,\\
&\langle m_2|v_1\rangle=0\,,\qquad
 \langle m_2|v_2\rangle=p\,,\qquad
 \langle m_2|v_3\rangle=q\,,\\
&\langle m_3|v_1\rangle=p\,,\qquad
 \langle m_3|v_2\rangle=0\,,\qquad
 \langle m_3|v_3\rangle=p-q\,,\\
&\langle m_4|v_1\rangle=0\,,\qquad
 \langle m_4|v_2\rangle=0\,,\qquad
 \langle m_4|v_3\rangle=m\,.
\end{align}
They define the monomials

\begin{equation}
X_b=z_2^p z_3^q\,,\qquad Y_b=z_1^p z_3^{p-q}\,,\qquad Z_b=z_1z_2z_3\,,\qquad S_b=z_3^m\,,
\end{equation}
which obey

\begin{equation}
X_bY_b=Z_b^p\,.
\end{equation}
Thus the lower-edge patch has the expected local form $\mathbb{C}^2/\mathbb{Z}_p\times\mathbb{C}^{\star}$.

\item \textbf{Left edge}: The left edge is generated by $\{v_1,v_3\}$. In this case

\begin{equation}
V=(\alpha,\beta)=(c,qa+mb+c)\,.
\end{equation}

Let

\begin{equation}
d={\rm gcd}(q,m)\,,\qquad q=dq'\,,\qquad m=dm'\,,
\end{equation}
so that ${\rm gcd}(q',m')=1$ and $\alpha-\beta\in d\mathbb{Z}$. Choosing a B\'ezout pair $(r,s)$ satisfying

\begin{equation}
q'r+m's=1\,,
\end{equation}
a convenient set of generators is

\begin{equation}
m_1=(-r,-s,d)\,,\qquad m_2=(r,s,0)\,,\qquad m_3=(0,0,1)\,,\qquad m_4=(m',-q',0)\,,
\end{equation}
where $m_4$ generates the $\mathbb{C}^{\star}$ direction. Their pairings with the toric vectors are

\begin{align}
&\langle m_1|v_1\rangle=d\,,\qquad
 \langle m_1|v_2\rangle=d-pr\,,\qquad
 \langle m_1|v_3\rangle=0\,,\\
&\langle m_2|v_1\rangle=0\,,\qquad
 \langle m_2|v_2\rangle=pr\,,\qquad
 \langle m_2|v_3\rangle=d\,,\\
&\langle m_3|v_1\rangle=1\,,\qquad
 \langle m_3|v_2\rangle=1\,,\qquad
 \langle m_3|v_3\rangle=1\,,\\
&\langle m_4|v_1\rangle=0\,,\qquad
 \langle m_4|v_2\rangle=m'p\,,\qquad
 \langle m_4|v_3\rangle=0\,.
\end{align}
The corresponding monomials are

\begin{equation}
X_l=z_1^d z_2^{d-pr}\,,\qquad
Y_l=z_2^{pr}z_3^d\,,\qquad
Z_l=z_1z_2z_3\,,\qquad
S_l=z_2^{\frac{mp}{d}}\,,
\end{equation}
and satisfy

\begin{equation}
X_lY_l=Z_l^d\,.
\end{equation}
Hence the left-edge patch is locally $\mathbb{C}^2/\mathbb{Z}_d\times\mathbb{C}^{\star}$.

\item \textbf{Right edge}: Finally, the right edge is generated by $\{v_2,v_3\}$, for which

\begin{equation}
V=(\alpha,\beta)=(pa+c,qa+mb+c)\,.
\end{equation}

Introducing
\begin{equation} 
e={\rm gcd}(p-q,m)\,,\qquad p-q=e(p-q)''\,,\qquad m=em''\,,
\end{equation}
we have ${\rm gcd}((p-q)'',m'')=1$ and $\alpha-\beta\in e\mathbb{Z}$. Let $(u,v)$ be a B\'ezout pair satisfying

\begin{equation}
(p-q)''u+m''v=1\,.
\end{equation}
A convenient choice of generators is then

\begin{equation}
m_1=(u,-v,e-pu)\,,\qquad m_2=(-u,v,pu)\,,\qquad m_3=(0,0,1)\,,\qquad m_4=(-m'',-(p-q)'',pm'')\,.
\end{equation}
Their pairings are

\begin{align}
&\langle m_1|v_1\rangle=e-pu\,,\qquad
 \langle m_1|v_2\rangle=e\,,\qquad
 \langle m_1|v_3\rangle=0\,,\\
&\langle m_2|v_1\rangle=pu\,,\qquad
 \langle m_2|v_2\rangle=0\,,\qquad
 \langle m_2|v_3\rangle=e\,,\\
&\langle m_3|v_1\rangle=1\,,\qquad
 \langle m_3|v_2\rangle=1\,,\qquad
 \langle m_3|v_3\rangle=1\,,\\
&\langle m_4|v_1\rangle=m''p\,,\qquad
 \langle m_4|v_2\rangle=0\,,\qquad
 \langle m_4|v_3\rangle=0\,.
\end{align}
This gives

\begin{equation}
X_r=z_1^{e-pu}z_2^e\,,\qquad Y_r=z_1^{pu}z_3^e\,,\qquad Z_r=z_1z_2z_3\,,\qquad S_r=z_1^{\frac{mp}{e}}\,,
\end{equation}
with

\begin{equation}
X_rY_r=Z_r^e\,.
\end{equation}
Thus the right-edge patch is locally $\mathbb{C}^2/\mathbb{Z}_e\times\mathbb{C}^{\star}$.

\end{itemize}

We now turn to the transition functions. Since all coordinates above are monomials in the same homogeneous variables $z_i$, the maps between overlapping patches follow directly by eliminating the $z_i$. For later reference, we collect the three resulting transformations:

\begin{itemize}

\item \textbf{Bottom--left}:

\begin{equation}
X_b=Y_l^{\frac{q}{d}}S_l^s\,,\qquad
Y_b=\frac{X_l^{\frac{p}{d}}Y_l^{\frac{p-q}{d}}}{S_l^s}\,,\qquad
Z_b=Z_l\,,\qquad
S_b=\frac{Y_l^{\frac{m}{d}}}{S_l^r}\,.
\end{equation}
Since $S_b$ and $S_l$ are non-vanishing on the overlap, this relation implies $Y_l\neq0$ there.

\item \textbf{Bottom--right}:
\begin{equation}
X_b=\frac{X_r^{\frac{p}{e}}Y_r^{\frac{q}{e}}}{S_r^v}\,,\qquad
Y_b=Y_r^{\frac{p-q}{e}}S_r^v\,,\qquad
Z_b=Z_r\,,\qquad
S_b=\frac{Y_r^{\frac{m}{e}}}{S_r^u}\,.
\end{equation}
Similarly, $S_b,S_r\neq0$ on the overlap implies $Y_r\neq0$.

\item \textbf{Left--right}: Defining

\begin{equation}
A=-(p-q)''r+m''s\,,\qquad
B=-vr-us\,,
\end{equation}
and

\begin{equation}
k=\frac{pm}{de}=(p-q)''m'+m''q'\,,\qquad
L=uq'-vm'\,,
\end{equation}
the transition functions take the form

\begin{equation}
Y_l=Y_r^A S_r^B Z_r^{pr}\,,\qquad
X_l=\frac{Z_r^d}{Y_r^A S_r^B Z_r^{pr}}\,,\qquad
Z_l=Z_r\,,\qquad
S_l=Y_r^{-k}S_r^L Z_r^{pm'}\,.
\end{equation}

\end{itemize}

\section{All-orders analysis of the gluing map} \label{app:allorders}

Recall that the equation induced on the left patch is
\begin{equation}
X_l^pY_l^p=Z_l^p+t\,C\,, \qquad C=\frac{Y_l^m}{S_l^r}\,.
\end{equation}
We now look for an invertible change of coordinates, depending only on the holomorphic coordinates and connected continuously to the identity, of the form
\begin{equation}
\label{eq:genericchange}
X_l\rightarrow X_l+A\,, \qquad Z_l\rightarrow Z_l+B\,, 
\end{equation}
where $A$ and $B$ are formal power series in $t$,
\begin{equation}
A=tA_1+t^2A_2+\cdots\,, \qquad B=tB_1+t^2B_2+\cdots\,.
\end{equation}
These functions are to be determined order by order by requiring the induced equation be mapped to
\begin{equation}
X_l^pY_l^p=Z_l^p\,.
\end{equation}
In the main text we have done this explicitly for the first few orders, obtaining the condition $m\geq p$. Let us now prove that it holds in general. To this end, introduce

\begin{equation}
U=X_lY_l\,,\qquad Z=Z_l\,,\qquad a=AY_l\,.
\end{equation}
Keeping $Y_l$ and $S_l$ fixed, the coordinate redefinition considered in section 3.1.2 is

\begin{equation}
(U,Z) \longrightarrow (\mathcal{U},\mathcal{Z})=(U+a\,, Z+a)\,.
\end{equation}
Hence, the equation $X_l^pY_l^p-Z_l^p-t\,C=U^p-Z^p-t\,C=0$ should become $\mathcal{U}^p-\mathcal{Z}^p=0$. We can therefore write

\begin{equation}
\label{eq:allorders}
U^p-Z^p-tC=(U+a)^p-(Z+a)^p\,, \qquad C=\frac{Y_l^m}{S_l^r}\,,
\end{equation}
Note that this equation is homogeneous under

\begin{equation}
\label{eq:homogeneity}
(U,Z,a,C)\longrightarrow(\lambda U,\lambda Z,\lambda a,\lambda^p C)\,.
\end{equation}

Let us now expand $a$ in a power series in the deformation parameter $t$

\begin{equation}
a=\sum_{n\geq1}t^n a_n\,.
\end{equation}
Expanding \eqref{eq:allorders} in powers of $a$, the equation takes the schematic form

\begin{equation}
D_1 a+D_2 a^2+\cdots+D_p a^p=- tC\,,
\end{equation}
where $D_k=\binom{p}{k}(U^{p-k}-Z^{p-k})$ depends only on $U$ and $Z$. At first order in $t$ one has $D_1a_1=- C$, and hence $a_1=Cf_1(U,Z)$. This structure persists recursively. Indeed, suppose that $a_n=C^n f_n(U,Z)$ for all $n<j$. At order $t^j$, every nonlinear contribution is a product $a_{j_1}\cdots a_{j_k}$ with $j_1+\cdots+j_k=j$, and is therefore proportional to $C^j$. The equation at this order consequently takes the form

\begin{equation}
D_1a_j+C^jF_j(U,Z)=0\,,
\end{equation}
for some function $F_j(U,Z)$. It follows by induction that

\begin{equation}
a_n=C^n f_n(U,Z)
\end{equation}
at every order. Moreover, since homogeneity implies that every coefficient $a_n$ has weight one under \eqref{eq:homogeneity}, one finds

\begin{equation}
f_n(\lambda U,\lambda Z)=\lambda^{1-pn}f_n(U,Z)\,.
\label{eq:fnscaling}
\end{equation}

Let us now approach the locus at which the extension of the gluing map has to be tested. As in section 3.1.2, write

\begin{equation}
U=Z+\epsilon\,.
\end{equation}
On the deformed geometry, when $Z\rightarrow0$,

\begin{equation}
\epsilon^p\sim C\sim Y_l^m\,, \qquad\Longrightarrow\qquad U,Z,\epsilon\sim Y_l^{m/p}\,.
\label{eq:allordersscaling}
\end{equation}
Hence, using \eqref{eq:fnscaling} and \eqref{eq:allordersscaling}, every coefficient scales as

\begin{equation}
a_n\sim C^n\left(Y_l^{m/p}\right)^{1-pn} \sim Y_l^{m/p}\,.
\end{equation}
Recalling that $a_n=A_nY_l$, we obtain

\begin{equation}
A_n\sim Y_l^{m/p-1}\,, \qquad B_n=A_nY_l\sim Y_l^{m/p}\,,
\qquad \forall\,n\geq1\,.
\label{eq:allordersresult}
\end{equation}
Thus the potentially singular shift has the same scaling at every order in $t$. Requiring the shift to remain finite as $Y_l\to0$ gives

\begin{equation}
m\geq p\,.
\end{equation}
Hence the scaling condition obtained explicitly at first and second order in section 3.1.2 persists to all orders for this local formal representative of the gluing map.

\section{The Q-Gorenstein smoothing of the T-cone}\label{app:qgorenstein}

In section \ref{sec:Tcones} we saw that, when the s-rule is saturated, $m=p$ and the deformation of the lower-edge patch extends to the global equation
\begin{equation}
AB-C^p-t=0\,.
\label{eq:QGsmoothing}
\end{equation}
We now show explicitly that this deformation is a Q-Gorenstein smoothing of the T-cone (in fact Gorenstein in this case).

For $m=p$, the T-cone is the orbifold $\mathbb{C}^3/\mathbb{Z}_{p^2}$ with generator $g$ acting as
\begin{equation}
(z_1,z_2,z_3)\longrightarrow \left(\xi^{-2p+q}z_1,\xi^{p-q}z_2,\xi^p z_3\right)\,, \qquad \xi^{p^2}=1\,.
\label{eq:TconeactionQG}
\end{equation}
Consider first the subgroup $H=\langle g^p\rangle\simeq\mathbb{Z}_p$, and define $\omega=\xi^p$. Its action is
\begin{equation}
(z_1,z_2,z_3)\longrightarrow \left(\omega^q z_1,\omega^{-q}z_2,z_3\right)\,.
\end{equation}
At saturation $m=p$, the standing assumption $e=\gcd(p-q,m)=1$ becomes $\gcd(p-q,p)=1$, equivalently $\gcd(p,q)=1$. Thus this is the standard $\mathbb{Z}_p$ action of type $(1,-1,0)$ up to a choice of generator. The invariant coordinates
\begin{equation}
A=z_2^p\,,\qquad B=z_1^p\,,\qquad C=z_1z_2\,,\qquad D=z_3
\end{equation}
obey
\begin{equation}
AB=C^p\,.
\end{equation}
Thus
\begin{equation}
\mathbb{C}^3/H\simeq
\left\{AB-C^p=0\right\}\times\mathbb{C}_D\,.
\end{equation}

It remains to quotient by the residual group $\mathbb{Z}_{p^2}/H\simeq\mathbb{Z}_p$. The action induced by $g$ on the invariant coordinates is
\begin{equation}
(A,B,C,D)\longrightarrow \left(\omega^{-q}A,\omega^qB,\omega^{-1}C,\omega D\right)\,.
\label{eq:residualaction}
\end{equation}
Indeed, for example $A=z_2^p\mapsto\xi^{p(p-q)}A=\omega^{-q}A$, while the remaining transformations follow analogously. We may therefore write the T-cone as
\begin{equation}
X_0=\frac{\left\{AB-C^p=0\right\}\times\mathbb{C}_D}{\mathbb{Z}_p}\,,
\end{equation}
with the residual $\mathbb{Z}_p$ acting with weights $(-q,q,-1,1)$ on $(A,B,C,D)$.

The deformation \eqref{eq:QGsmoothing} is invariant under \eqref{eq:residualaction}, since both $AB$ and $C^p$ are invariant. It therefore defines the family
\begin{equation}
X_t=\frac{\left\{AB-C^p=t\right\}\times\mathbb{C}_D}{\mathbb{Z}_p}\,.
\label{eq:QGfamily}
\end{equation}
For $t\neq0$, the hypersurface upstairs is smooth. Indeed, a singular point of $AB-C^p=t$ would require
\begin{equation}
A=B=C=0\,,
\end{equation}
which is incompatible with $t\neq0$. Moreover, the residual $\mathbb{Z}_p$ acts freely on this generic fiber. To see this, consider a non-trivial element $\omega^k$. Since the weights of $C$ and $D$ are $-1$ and $1$, respectively, a fixed point must have $C=D=0$. Since $\gcd(p,q)=1$, the weights of $A$ and $B$ are also primitive, and hence a fixed point must furthermore have $A=B=0$. Again this is incompatible with $AB-C^p=t\neq0$. It follows that $X_t$ is smooth for $t\neq0$.

Finally, the family is Q-Gorenstein. Before taking the residual quotient, the total space is a hypersurface, and hence its relative canonical bundle is Cartier. A local holomorphic three-form on a fiber, for instance on the patch $A\neq0$, is
\begin{equation}
\Omega=\frac{dA\wedge dC\wedge dD}{A}\,.
\end{equation}
Under \eqref{eq:residualaction}, the numerator and denominator acquire the same factor $\omega^{-q}$, so that
\begin{equation}
\Omega\longrightarrow\Omega\,.
\end{equation}
The holomorphic volume form therefore descends through the residual quotient. Hence, the relative canonical divisor of the quotient family is Cartier, and in particular Q-Cartier.

We have thus exhibited a flat one-parameter family whose central fiber is the T-cone, whose generic fiber is smooth, and whose relative canonical divisor is Q-Cartier. Hence \eqref{eq:QGfamily} is a Q-Gorenstein smoothing of the T-cone. The deformation induced geometrically by ending the $p$ external 5-branes on a common 7-brane is therefore precisely a Q-Gorenstein smoothing in the saturating case $m=p$.

%======================================================
\bibliographystyle{JHEP}
\bibliography{mybib}

\providecommand{\href}[2]{#2}\begingroup\raggedright\begin{thebibliography}{10}

\bibitem{Morrison:1996xf}
D.~R. Morrison and N.~Seiberg, \emph{{Extremal transitions and five-dimensional
  supersymmetric field theories}},
  \href{http://dx.doi.org/10.1016/S0550-3213(96)00592-5}{\emph{Nucl. Phys. B}
  {\bf 483} (1997) 229--247}, [\href{http://arxiv.org/abs/hep-th/9609070}{{\tt
  hep-th/9609070}}].

\bibitem{Douglas:1996xp}
M.~R. Douglas, S.~H. Katz and C.~Vafa, \emph{{Small instantons, Del Pezzo
  surfaces and type I-prime theory}},
  \href{http://dx.doi.org/10.1016/S0550-3213(97)00281-2}{\emph{Nucl. Phys. B}
  {\bf 497} (1997) 155--172}, [\href{http://arxiv.org/abs/hep-th/9609071}{{\tt
  hep-th/9609071}}].

\bibitem{Intriligator:1997pq}
K.~A. Intriligator, D.~R. Morrison and N.~Seiberg, \emph{{Five-dimensional
  supersymmetric gauge theories and degenerations of Calabi-Yau spaces}},
  \href{http://dx.doi.org/10.1016/S0550-3213(97)00279-4}{\emph{Nucl. Phys. B}
  {\bf 497} (1997) 56--100}, [\href{http://arxiv.org/abs/hep-th/9702198}{{\tt
  hep-th/9702198}}].

\bibitem{Closset:2018bjz}
C.~Closset, M.~Del~Zotto and V.~Saxena, \emph{{Five-dimensional SCFTs and gauge
  theory phases: an M-theory/type IIA perspective}},
  \href{http://dx.doi.org/10.21468/SciPostPhys.6.5.052}{\emph{SciPost Phys.}
  {\bf 6} (2019) 052}, [\href{http://arxiv.org/abs/1812.10451}{{\tt
  1812.10451}}].

\bibitem{Aharony:1997bh}
O.~Aharony, A.~Hanany and B.~Kol, \emph{{Webs of (p,q) five-branes,
  five-dimensional field theories and grid diagrams}},
  \href{http://dx.doi.org/10.1088/1126-6708/1998/01/002}{\emph{JHEP} {\bf 01}
  (1998) 002}, [\href{http://arxiv.org/abs/hep-th/9710116}{{\tt
  hep-th/9710116}}].

\bibitem{Leung:1997tw}
N.~C. Leung and C.~Vafa, \emph{{Branes and toric geometry}},
  \href{http://dx.doi.org/10.4310/ATMP.1998.v2.n1.a4}{\emph{Adv. Theor. Math.
  Phys.} {\bf 2} (1998) 91--118},
  [\href{http://arxiv.org/abs/hep-th/9711013}{{\tt hep-th/9711013}}].

\bibitem{Benini:2009gi}
F.~Benini, S.~Benvenuti and Y.~Tachikawa, \emph{{Webs of five-branes and N=2
  superconformal field theories}},
  \href{http://dx.doi.org/10.1088/1126-6708/2009/09/052}{\emph{JHEP} {\bf 09}
  (2009) 052}, [\href{http://arxiv.org/abs/0906.0359}{{\tt 0906.0359}}].

\bibitem{Bourget:2023wlb}
A.~Bourget, A.~Collinucci and S.~Schafer-Nameki, \emph{{Generalized toric
  polygons, T-branes, and 5d SCFTs}},
  \href{http://dx.doi.org/10.21468/SciPostPhys.18.3.079}{\emph{SciPost Phys.}
  {\bf 18} (2025) 079}, [\href{http://arxiv.org/abs/2301.05239}{{\tt
  2301.05239}}].

\bibitem{Hanany:1996ie}
A.~Hanany and E.~Witten, \emph{{Type IIB superstrings, BPS monopoles, and
  three-dimensional gauge dynamics}},
  \href{http://dx.doi.org/10.1016/S0550-3213(97)00157-0}{\emph{Nucl. Phys. B}
  {\bf 492} (1997) 152--190}, [\href{http://arxiv.org/abs/hep-th/9611230}{{\tt
  hep-th/9611230}}].

\bibitem{Bachas:1997sc}
C.~Bachas and M.~B. Green, \emph{{A Classical manifestation of the Pauli
  exclusion principle}},
  \href{http://dx.doi.org/10.1088/1126-6708/1998/01/015}{\emph{JHEP} {\bf 01}
  (1998) 015}, [\href{http://arxiv.org/abs/hep-th/9712187}{{\tt
  hep-th/9712187}}].

\bibitem{Kitao:1998mf}
T.~Kitao, K.~Ohta and N.~Ohta, \emph{{Three-dimensional gauge dynamics from
  brane configurations with (p,q)-fivebrane}},
  \href{http://dx.doi.org/10.1016/S0550-3213(98)00726-3}{\emph{Nucl. Phys. B}
  {\bf 539} (1999) 79--106}, [\href{http://arxiv.org/abs/hep-th/9808111}{{\tt
  hep-th/9808111}}].

\bibitem{Mikhailov:1998bx}
A.~Mikhailov, N.~Nekrasov and S.~Sethi, \emph{{Geometric realizations of BPS
  states in N=2 theories}},
  \href{http://dx.doi.org/10.1016/S0550-3213(98)80001-1}{\emph{Nucl. Phys. B}
  {\bf 531} (1998) 345--362}, [\href{http://arxiv.org/abs/hep-th/9803142}{{\tt
  hep-th/9803142}}].

\bibitem{Bergman:1998ej}
O.~Bergman and A.~Fayyazuddin, \emph{{String junction transitions in the moduli
  space of N=2 SYM}},
  \href{http://dx.doi.org/10.1016/S0550-3213(98)00553-7}{\emph{Nucl. Phys. B}
  {\bf 535} (1998) 139--151}, [\href{http://arxiv.org/abs/hep-th/9806011}{{\tt
  hep-th/9806011}}].

\bibitem{DeWolfe:1998bi}
O.~DeWolfe, T.~Hauer, A.~Iqbal and B.~Zwiebach, \emph{{Constraints on the BPS
  spectrum of N=2, D = 4 theories with A-D-E flavor symmetry}},
  \href{http://dx.doi.org/10.1016/S0550-3213(98)00652-X}{\emph{Nucl. Phys. B}
  {\bf 534} (1998) 261--274}, [\href{http://arxiv.org/abs/hep-th/9805220}{{\tt
  hep-th/9805220}}].

\bibitem{Iqbal:1998xb}
A.~Iqbal, \emph{{Selfintersection number of BPS junctions in backgrounds of
  three-branes and seven-branes}},
  \href{http://dx.doi.org/10.1088/1126-6708/1999/10/032}{\emph{JHEP} {\bf 10}
  (1999) 032}, [\href{http://arxiv.org/abs/hep-th/9807117}{{\tt
  hep-th/9807117}}].

\bibitem{Alexeev:2024bko}
V.~Alexeev, H.~Arg{\"u}z and P.~Bousseau, \emph{{Non-toric Brane Webs,
  Calabi{\textendash}Yau 3-Folds, and 5d SCFTs}},
  \href{http://dx.doi.org/10.1007/s00220-025-05461-9}{\emph{Commun. Math.
  Phys.} {\bf 406} (2025) 287}, [\href{http://arxiv.org/abs/2410.04714}{{\tt
  2410.04714}}].

\bibitem{Franco:2023flw}
S.~Franco and R.-K. Seong, \emph{{Twin theories, polytope mutations and quivers
  for GTPs}}, \href{http://dx.doi.org/10.1007/JHEP07(2023)034}{\emph{JHEP} {\bf
  07} (2023) 034}, [\href{http://arxiv.org/abs/2302.10951}{{\tt 2302.10951}}].

\bibitem{Arias-Tamargo:2024fjt}
G.~Arias-Tamargo, S.~Franco and D.~Rodr\'\i{}guez-G\'omez, \emph{{The geometry
  of GTPs and 5d SCFTs}},
  \href{http://dx.doi.org/10.1007/JHEP07(2024)159}{\emph{JHEP} {\bf 07} (2024)
  159}, [\href{http://arxiv.org/abs/2403.09776}{{\tt 2403.09776}}].

\bibitem{Ilten_2012}
N.~O. Ilten, \emph{{Mutations of {Laurent} polynomials and flat families with
  toric fibers}},
  \href{http://dx.doi.org/10.3842/sigma.2012.047}{\emph{Symmetry, Integrability
  and Geometry: Methods and Applications} (July, 2012) }.

\bibitem{Collinucci:2026kom}
A.~Collinucci, \emph{{M-theory geometries from five-brane webs, seven-branes,
  and T-branes}},  \href{http://arxiv.org/abs/2607.05588}{{\tt 2607.05588}}.

\bibitem{Bergman:2020myx}
O.~Bergman and D.~Rodr\'\i{}guez-G\'omez, \emph{{The Cat\textquoteright{}s
  Cradle: deforming the higher rank E$_{1}$ and $ {\tilde{E}}_1 $ theories}},
  \href{http://dx.doi.org/10.1007/JHEP02(2021)122}{\emph{JHEP} {\bf 02} (2021)
  122}, [\href{http://arxiv.org/abs/2011.05125}{{\tt 2011.05125}}].

\bibitem{CarrenoBolla:2024fxy}
I.~Carre{\~n}o~Bolla, S.~Franco and D.~Rodr{\'\i}guez-G{\'o}mez, \emph{{The 5d
  tangram: brane webs, 7-branes and primitive T-cones}},
  \href{http://dx.doi.org/10.1007/JHEP05(2025)175}{\emph{JHEP} {\bf 05} (2025)
  175}, [\href{http://arxiv.org/abs/2411.01510}{{\tt 2411.01510}}].

\bibitem{Shepherd1988}
J.~Kollár and N.~I. Shepherd-Barron, \emph{Threefolds and deformations of
  surface singularities}, {\emph{Inventiones mathematicae} {\bf 91} (1988)
  299--338}.

\bibitem{Closset:2019juk}
C.~Closset and M.~Del~Zotto, \emph{{On 5D SCFTs and their BPS quivers. Part I:
  B-branes and brane tilings}},
  \href{http://dx.doi.org/10.4310/ATMP.2022.v26.n1.a2}{\emph{Adv. Theor. Math.
  Phys.} {\bf 26} (2022) 37--142}, [\href{http://arxiv.org/abs/1912.13502}{{\tt
  1912.13502}}].

\bibitem{Franco:2026bie}
S.~Franco and D.~Rodr{\'\i}guez-G{\'o}mez, \emph{{On the Origin of Toric
  Diagrams}},  \href{http://arxiv.org/abs/2607.19460}{{\tt 2607.19460}}.

\bibitem{Franco:2026hiz}
S.~Franco and D.~Rodr{\'\i}guez-G{\'o}mez, \emph{{Towards Generalized Dimers
  for GTPs: $\mathcal{N}=2$ Fractional Branes at Infinite Coupling}},
  \href{http://arxiv.org/abs/2607.20607}{{\tt 2607.20607}}.

\bibitem{CoxLittleSchenck}
D.~A. Cox, J.~B. Little and H.~K. Schenck,
  \emph{{\href{https://doi.org/10.1090/gsm/124}{Toric Varieties}}}, vol.~124 of
  \emph{Graduate Studies in Mathematics}.
\newblock American Mathematical Society, 2011.

\end{thebibliography}\endgroup
%======================================================

\end{document}